# Characterization of a novel time-resolved, real-time scintillation dosimetry system for ultra-high dose rate radiation therapy applications

Alexander Baikalov[1,2,3], Daline Tho[1], Kevin Liu[1,4], Stefan Bartzsch[2,3], Sam Beddar[1,4*], Emil Schüler[1,4*]

[1] Department of Radiation Physics, The University of Texas MD Anderson Cancer Center, Houston, TX, United States
[2] Department of Radiation Oncology, School of Medicine and Klinikum rechts der Isar, Technical University of Munich, Munich, Germany
[3] Institute of Radiation Medicine, Helmholtz Zentrum München GmbH, German Research Center for Environmental Health, Neuherberg, Germany
[4] The University of Texas MD Anderson Cancer Center UTHealth Graduate School of Biomedical Sciences, Houston, TX, United States

*Correspondence
Emil Schüler: eschueler@mdanderson.org
Sam Beddar: abeddar@mdanderson.org

Funding
Research reported in this publication was supported in part by the National Cancer Institute of the National Institutes of Health under Award Number R01CA266673, by the University Cancer Foundation via the Institutional Research Grant program at MD Anderson Cancer Center, by a grant from MD Anderson's Division of Radiation Oncology, by Cancer Center Support Grant P30 CA016672 from the National Cancer Institute of the National Institutes of Health to The University of Texas MD Anderson Cancer Center, by a UTHealth Innovation for Cancer Prevention Research Training Program Pre-doctoral Fellowship (Cancer Prevention and Research Institute of Texas grant #RP210042), and by the Klinikum rechts der Isar of the Technical University of Munich. The content is solely the responsibility of the authors and does not necessarily represent the official views of the National Institutes of Health or the Cancer Prevention and Research Institute of Texas.

Data Sharing
The raw data supporting the conclusion of this article will be made available by the authors, without undue reservation.

Acknowledgements
The authors thank François Therriault-Proulx and Benjamin Côté (MedScint, Quebec City, Canada) for their technical advice while performing this work. We would like to also thank Amy Ninetto of the Research Medical Library at MD Anderson Cancer Center for editing the manuscript.

# Abstract

**Background:** Scintillation dosimetry has promising qualities for ultra-high dose rate (UHDR) radiotherapy (RT), but no system has shown compatibility with mean dose rates ($\overline{DR}$) above 100 Gy/s and doses per pulse ($D_p$) exceeding 1.5 Gy typical of UHDR (FLASH)-RT. The aim of this study was to characterize a novel scintillator dosimetry system with the potential of accommodating UHDRs.
**Methods and Materials:** A thorough dosimetric characterization of the system was performed on an UHDR electron beamline. The system's response as a function of dose, $\overline{DR}$, $D_p$, and the pulse dose rate $DR_p$ was investigated, together with the system's dose sensitivity (signal per unit dose) as a function of dose history. The capabilities of the system for time-resolved dosimetric readout were also evaluated.
**Results:** Within a tolerance of ±3%, the system exhibited dose linearity and was independent of $\overline{DR}$ and $D_p$ within the tested ranges of 1.8–1341 Gy/s and 0.005–7.68 Gy, respectively. A 6% reduction in the signal per unit dose was observed as $DR_p$ was increased from 8.9e4–1.8e6 Gy/s. Additionally, the dose delivered per integration window of the continuously sampling photodetector had to remain between 0.028 and 11.64 Gy to preserve a stable signal response per unit dose. The system accurately measured $D_p$ of individual pulses delivered at up to 120 Hz. The day-to-day variation of the signal per unit dose at a reference setup varied by up to ±13% but remained consistent (<±2%) within each day of measurements and showed no signal loss as a function of dose history.
**Conclusions:** With daily calibrations and $DR_p$ specific correction factors, the system reliably provides real-time, millisecond-resolved dosimetric measurements of pulsed conventional and UHDR beams from typical electron linacs, marking an important advancement in UHDR dosimetry and offering diverse applications to FLASH-RT and related fields.

# Introduction

Precise and reliable dosimetry is a fundamental component of safe and successful radiation therapy (RT). Recent advances in ultra-high dose rate (UHDR, higher than ~40 Gy/s) FLASH-RT protocols, in contrast with conventional dose rate (CDR, ~0.1Gy/s) RT, present unique challenges to dosimetry. Saturation effects due to the high particle fluxes present at UHDRs render most conventional radiation detectors unreliable, necessitating the development of specifically designed UHDR detectors [1].

The essential characteristics of traditional detectors for CDR-RT, including real-time signal readout, high accuracy and precision, a linear response to dose, and independence from beam quality and dose rate, continue to be crucial for UHDR detectors. However, UHDR detectors face significantly greater demands in these aspects, particularly for high-energy electron and photon deliveries, where conventional detectors and dosimeters display signal saturation and dose rate–dependent readouts [1-3]. These deliveries, using UHDR-capable linear accelerators, often comprise just one or a few microsecond-long pulses at up to 360 Hz, with doses per pulse ($D_p$) up to ~10 Gy, pulse dose rates ($DR_p$) on the order of MGy/s, and mean dose rates ($\overline{DR}$) on the order of kGy/s [4]. UHDR detectors must therefore be dose rate independent over an extreme range of dose rates and exhibit dose linearity across a large range of nearly instantaneously delivered doses. Ideally, a UHDR detector should also have a high enough temporal resolution to differentiate between pulses, with a goal of sub-microsecond resolution to measure parameters like pulse width (*PW*) [5].

Scintillation dosimetry has been studied extensively in various CDR-RT contexts, and scintillator detectors have many characteristics that are ideally suited for FLASH-RT applications [6]. Organic plastic scintillators have a very fast response time (<15 ns) with a linear dose response, are water equivalent at relevant energies, are dose-rate independent (at CDRs), and can be made very small whilst retaining sensitivity [7-10]. Plastic scintillators operate on the following principle: radiation-induced electronic excitation of the scintillating material results in photon emission following deexcitation (within nanoseconds) directly proportional to the absorbed dose. An optical fiber is typically used to guide this scintillation signal to a detector. However, Cerenkov and fluorescence radiation from within both the scintillator and the fiber contaminate the scintillation signal and must be dealt with appropriately as they are not dose-proportional; many methods for this have been developed [9, 11, 12]. Additionally, quenching effects due to partially non-radiative relaxation after high linear energy transfer (LET) radiation must be considered [13, 14].

Plastic scintillators thus appear to be good candidates for low-LET UHDR beamlines, though limited research into their responses at UHDRs exists. One plastic scintillator was studied under x-ray radiation, indicating good performance up to the highest tested dose rate of 118.0 Gy/s [15]. A 2D plastic scintillation detector[16] and three point detectors[17-19] were studied under UHDR electron radiation, also indicating good performance at the lower end ($D_p$ < 1.5 Gy and $\overline{DR}$ < 380 Gy/s) of UHDR parameter ranges. However, radiation damage was noted[18, 19] and, at more extreme values of $D_p$, $\overline{DR}$, and the pulse repetition frequency (*PRF*), nonlinear responses and signal saturation were observed [19].

In this work, we performed a detailed characterization of a novel FLASH-dedicated scintillation dosimetry system and tested its capabilities in providing real-time, highly time-resolved dosimetric data. We demonstrated its dose linearity and pulse-by-pulse dose measurement capabilities up to the highest tested values of *PRF*, $\overline{DR}$, and $D_p$ of 120 Hz, 1340 Gy/s, and 7.7 Gy, respectively.

# Methods and Material

All parameter symbols used in the present work are listed in **Figure 1a**.

**Scintillation Dosimetry System**
The prototype Hyperscint RP-FLASH scintillation dosimetry system (MedScint, Quebec City, Canada) comprises a plastic scintillator probe with a cylindrical active volume of 1 mm diameter x 3 mm length connected via a polymethyl methacrylate plastic optical fiber to a spectrometer with a cooled 2D photodetector array. During measurement, the photodetector collects the light spectrum from the probe over a set 'integration window' (*IW*) after which an automatic signal readout process is performed. The integration window determines the sampling frequency ($f_s = 1/IW$) of the measurement. If *IW* > 40 ms ($f_s$ < 25 Hz), the system operates in 'continuous mode', whereby it continues to record at the set sampling frequency until the measurement is stopped by the user. Otherwise, if $IW$ < 40 ms, the system operates in 'FLASH mode', where a fixed number of IWs (samples) are recorded, with a maximum sampling frequency of 1000 Hz. The recorded spectrum per IW is automatically processed by the vendor-supplied HyperDose software using a hyperspectral approach to isolate the scintillation, fluorescence, and Cerenkov signals [20]. For the reported measurements, the system was pre-calibrated by the manufacturer.

**Measurement Setup**

Irradiation measurements were performed with a 9 MeV electron beam from an electron linear accelerator (Mobetron, IntraOp, Sunnyvale, CA, USA) capable of both CDR and UHDR radiation delivery. For all measurements, the probe was placed between two 1-cm sheets of water-equivalent, flexible bolus material, with the active region of the probe centered in the radiation field **(Figure 1b)**. At least 7 cm of backscatter solid water material was placed underneath the bolus sheets.

| Symbol | Name | Unit | Note |
|---|---|---|---|
| $N_p$ | Number of pulses | | |
| $SSD$ | Source-to-surface distance | m | |
| $PRF$ | Pulse repetition frequency | Hz | |
| $PW$ | Pulse width | s | FWHM |
| $D$ | Dose | Gy | |
| $D_p$ | Dose per pulse | Gy | $D/N_p$ |
| $DR_p$ | Pulse dose rate | Gy/s | $D_p/PW$ |
| $\overline{DR}$ | Mean dose rate | Gy/s | $D/((N_p - 1)/PRF)$ |
| $IW$ | Integration window | s | of photodetector |
| $D_{iw}$ | Dose per integration window | Gy | |
| $f_s$ | Sampling frequency | Hz | $1/IW$ |

(a)

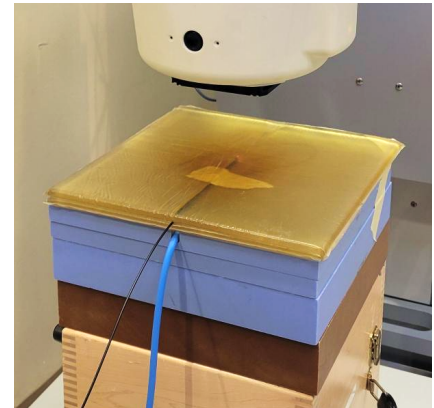

(b)

*Figure 1. (a) Parameter symbols used in this work. (b) Measurement setup of the scintillator probe under the Mobetron beamline. Tape was used to secure the probe in place between the two sheets of semi-transparent bolus material. The white treatment head of the Mobetron is visible at the top of the image. The blue cable from the ionization chamber (only included during conventional dose rate measurements), embedded in the 2 cm of solid water directly below the bolus material, is also visible.*

Before measurement, the spectrometer was left powered on for a few minutes until it reached a stable temperature, which was indicated by the system's software. Unless otherwise specified, each UHDR measurement comprised the delivery of 3 pulses at 30 Hz, whereby the average signal per pulse was recorded. Each 3-pulse measurement was performed in triplicate, the average of which is reported with an error bar representing one standard deviation. Some graphs contain error bars that are smaller than the symbols used to represent the values and thus are not visible.

$IW$ was set to 4.1 ms ($f_s$ = 244 Hz) to exceed the Nyquist frequency of the highest possible beam delivery $PRF$ of 120 Hz, and the number of samples per measurement was set to 800, resulting in a measurement length of 3.28 s. The start of each measurement was timed to coincide with the delivery of the beam.

To determine the dose delivered to the probe for each measurement, each setup was calibrated using dose rate–independent radiochromic film (Gafchromic EBT3, Ashland Inc., Covington, KY, USA) applying a previously described protocol [21]. For CDR measurements, the dose was monitored using an Advanced Markus ionization chamber (PTW-Freiburg GmbH, Freiburg, Germany) placed at a set location below the probe (embedded in the solid water **[Figure 1b]**). For UHDR measurements, inline beam current transformers (BCTs) (Bergoz Instrumentation, Saint-Genis-Pouilly, France) were used as previously described [22]. The BCTs yield highly time-resolved measurements of the beam current for each individual pulse delivered and were used as the reference against which the scintillator system's response was compared. Both the ion chamber (in CDR mode) and the BCTs (in UHDR mode) were calibrated to the film dose at the probe location. The statistical errors across triplicate film dose measurements were propagated to the final reported values.

**System Dose Response and Stability at CDRs**

All CDR measurements were performed using the same reference setup **(Figure 1b)** under an uncollimated field at a fixed source-to-surface distance ($SSD$) of 35.8 cm, varying only the parameters $D$, $IW$, and therefore $D_{iw}$. First, the system's response as a function of $D_{iw}$ was evaluated from 0.007–14 Gy by adjusting $D$ and $IW$. Second, the system's response as a function of $D$ was evaluated from 0.05–17.2 Gy at fixed values of $IW$ = 1s. $D_{iw}$ therefore ranged from 0.05–0.14 Gy. Finally, the stability of the system over 10 non-consecutive days and ~3 kGy of accumulated dose was evaluated by periodically measuring a triplicate delivery of 2.42 ± 0.04 Gy with $IW$ = 1 s (thus, $D_{iw}$ = 0.14).

*Specific parameters: $PW$ = 1.2 µs, $\overline{DR}$ = 0.14 Gy/s, $DR_p$ = 3.8e3 Gy/s, $D_p$ = 5 mGy.*

**System Dependency on $D_p$**

To study the system's response at UHDRs as a function of $D_p$, the $D_p$ was varied by (1) changing the $PW$ while keeping the $DR_p$ constant **(Figure 2a)**, or by (2) changing $DR_p$ while keeping the $PW$ constant **(Figure 2b)**.

In condition 1, the probe was exposed to pulses of varying $PW$ (0.5–4 µs) at a constant $DR_p$. This was repeated for two different $DR_p$, the highest and lowest possible with the experimental setup, to achieve a wider range of $D_p$.

*Specific parameters: $PW$ = 0.5–4 µs, $DR_p$ = 8.9e4 Gy/s; 1.9e6 Gy/s, $\overline{DR}$ = 1.8–15.8 Gy/s; 43–328 Gy/s, $D_p$ = 0.04–0.35 Gy; 0.95–7.28 Gy.*

In condition 2, to study the system's linearity with $D_p$ at a constant $PW$ but varying $DR_p$, the probe was irradiated at varying $SSD$. Since the field was uncollimated, the amount of exposed optical fiber increased with the $SSD$.

*Specific parameters: $PW$ = 4 µs, $PRF$ = 30 Hz, $DR_p$ = 8.9e4–1.9e6 Gy/s, $\overline{DR}$ =16–346 Gy/s, $D_p$ = 0.36–7.68 Gy, $SSD$ = 25.8–111.2 cm*

**System Dependency on $\overline{DR}$, PW, and $DR_p$**

To determine if the system's response was influenced by $\overline{DR}$, the *PRF* was varied while keeping all other parameters constant **(Figure 2c)**. The *PRF* was varied between 5–120 Hz, resulting in a total time between two sequential pulses of 8.3–200 ms. These measurements were repeated at two different *SSD*s, and thus two different $DR_p$, the highest and lowest possible with the experimental setup, to cover a wider range of $\overline{DR}$.

*Specific parameters: PW = 4 µs, PRF = 5–120 Hz, $DR_p$ = 8.6e4 Gy/s; 1.9e6 Gy/s, $\overline{DR}$ = 12–46 Gy/s; 57–1,341 Gy/s, $D_p$ = 0.34 Gy; 7.6 Gy.*

To determine the system's response when varying both *PW* and $DR_p$ at a constant $D_p$, the probe was exposed to the same $D_p$ by increasing the *PW* as the *SSD* was increased **(Figure 2d)**. These measurements were repeated for two different values of $D_p$: 4.01 ± 0.12 Gy and 1.00 ± 0.02 Gy.

*Specific parameters: PW = 0.5–4 µs; 2–4 µs; $DR_p$ = 2.5e5–1.8e6 Gy/s; 1e6–1.8e6 Gy/s, $\overline{DR}$ = 45 Gy/s; 180 Gy/s, $D_p$ = 1 Gy; 4 Gy.*

**Pulse Discrimination and Pulse-by-pulse $D_p$ Measurement**

To study the system's ability to differentiate between pulses and reliably measure $D_p$ of individual pulses, 300 pulses were delivered at 30 Hz and $D_p$ = 0.1 Gy. The system's response was recorded and compared with the beam current recorded by the BCTs for each individual pulse. This was performed at three different values of *IW* to vary $f_s$: once equal to the Nyquist frequency, once slightly greater than the Nyquist frequency as recommended by the manufacturer, and once at approximately double that frequency.

*Specific parameters: PW = 1 µs; PRF = 30 Hz, $DR_p$ = 1e5 Gy/s, $\overline{DR}$ = 2.8 Gy/s, $D_p$ = 0.1 Gy*

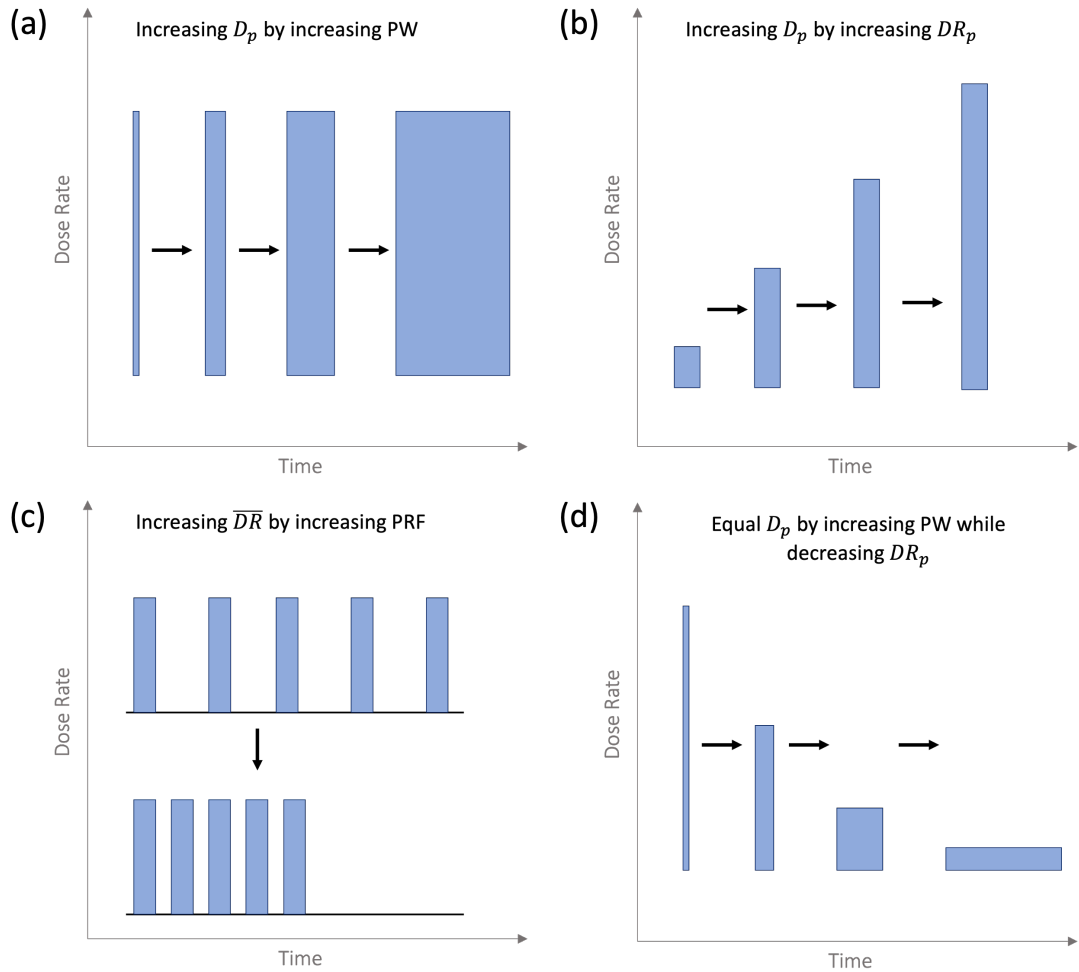

*Figure 2.* Schematic of the various pulsed beam parameter changes possible with the FLASH Mobetron. The black arrows indicate what was being compared in each experiment type.

# Results

**System Dose Response and Stability at CDRs**

The signal per unit dose varied by less than ±2% for $D_{iw}$ within 0.028–11.56 Gy but was lower when $D_{iw}$ < 0.007 Gy or $D_{iw}$ > 12.5 Gy **(Figure 3a)**. The signal increased linearly with $D$ across the entire tested range; the dose-normalized signal varied by less than ±3% **(Figure 3b)**.

Periodic measurements over the course of 10 non-consecutive days of measurement reveal a general variance of the signal by up to ±13%, during which the probe was exposed to ~3 kGy of accumulated dose **(Figure 3c)**. A variation of less than ±2% was observed within each day. No signal degradation as a function of either time or dose was evident. Immediately subsequent measurements within each triplicate varied on average by 0.2 ± 0.2 %.

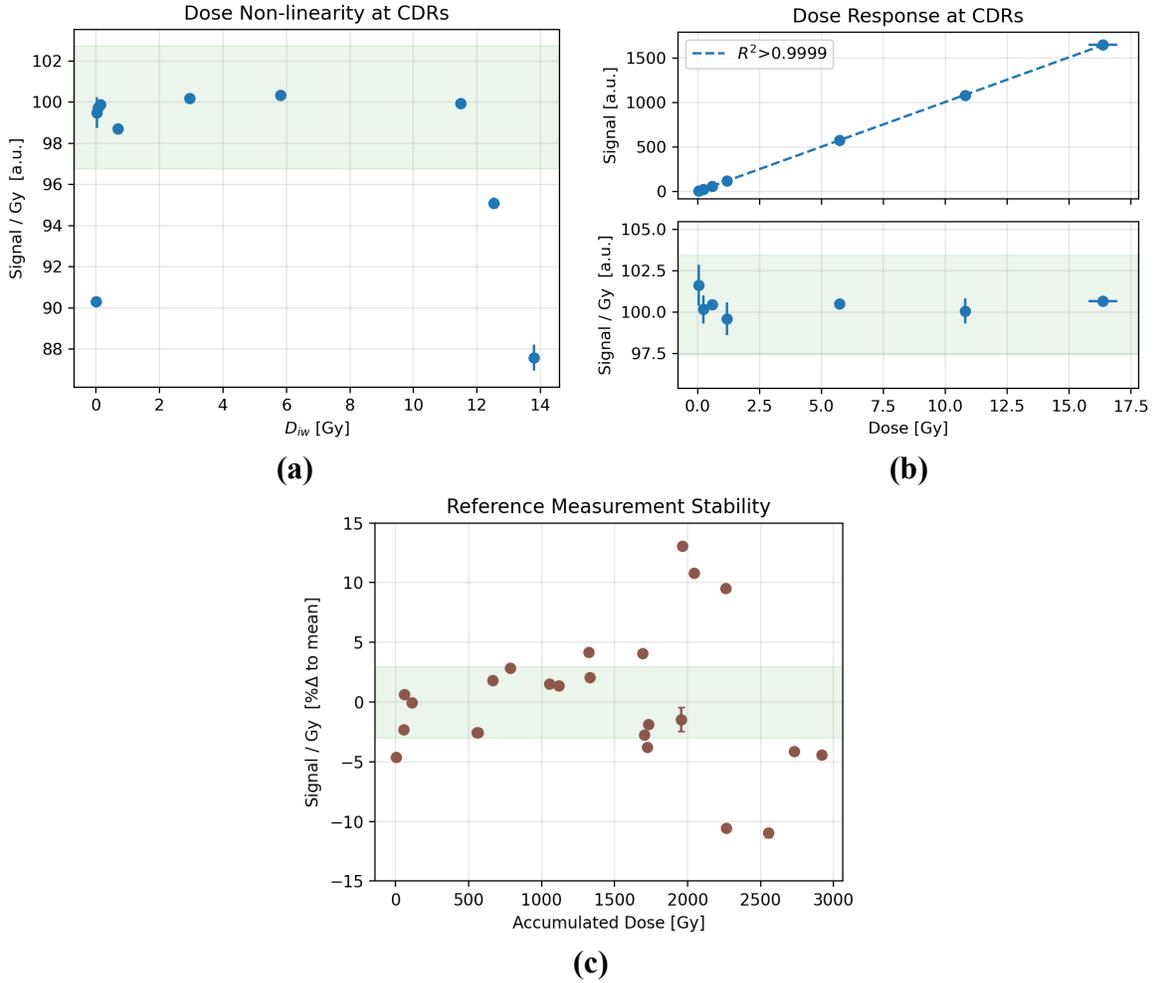

*Figure 3. (a) Signal per unit dose over a wide range of $D_{iw}$. (b) Dose response (top) and signal per unit dose (bottom) at CDRs wherein $D_{iw}$ was kept between 0.05 and 0.14 Gy. (c) Dose-normalized signal change, reported as a percent change relative to the mean, at a reference setup over 10 non-consecutive days of measurements and ~3 kGy of accumulated dose. The green shaded region indicates a ±3% variance from the mean.*

**System Dependency on $D_p$**

The system response was linear with dose in the range tested; the dose-normalized signal varied by less than ±3% **(Figure 4a-b)**. The signal also remained linear with dose as $D_p$ was changed from 0.36–7.68 Gy at a constant *PW* by varying the *SSD* **(Figure 4c)**. A ~6% decrease in the signal response per unit dose was observed as $D_p$ increased. This trend persisted even after a fresh recalibration of the system and after collimation of the field to equalize the amount of exposed optical fiber at each *SSD*.

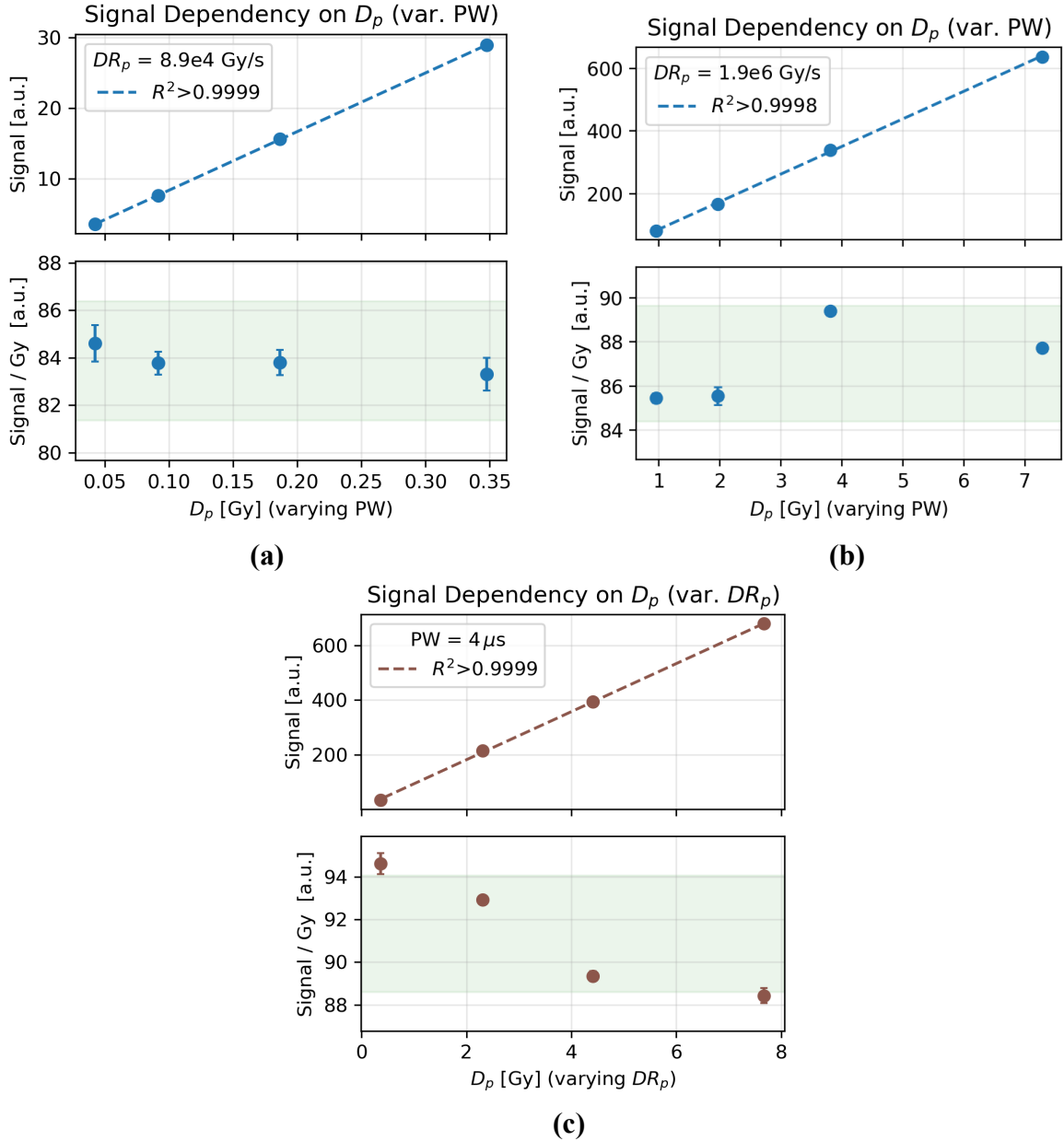

*Figure 4. Dose response as $D_p$ was increased by increasing PW at constant values of $DR_p$ of either **(a)** 8.9e4Gy/s or **(b)** 1.9e6Gy/s. **(c)** The signal increased linearly with $D_p$ as the $DR_p$ was increased at a constant PW. The signal per unit dose decreased with increasing $DR_p$. The green shaded region indicates a ±3% variance from the mean.*

**System Dependency on $\overline{DR}$, PW, and $DR_p$**

The signal varied by less than ±1% with changes in $\overline{DR}$ at both tested $DR_p$ (8.6e4 Gy/s and 1.9e6 Gy/s) **(Figure 5a)**. The signal per unit dose was unaffected by varying *PW* and *SSD* at a constant $D_p$ at both tested values of $D_p$ **(Figure 5b)**. Although the values varied by ±3%, no general trend is apparent, and the variance is comparable in magnitude to the uncertainty of each measurement.

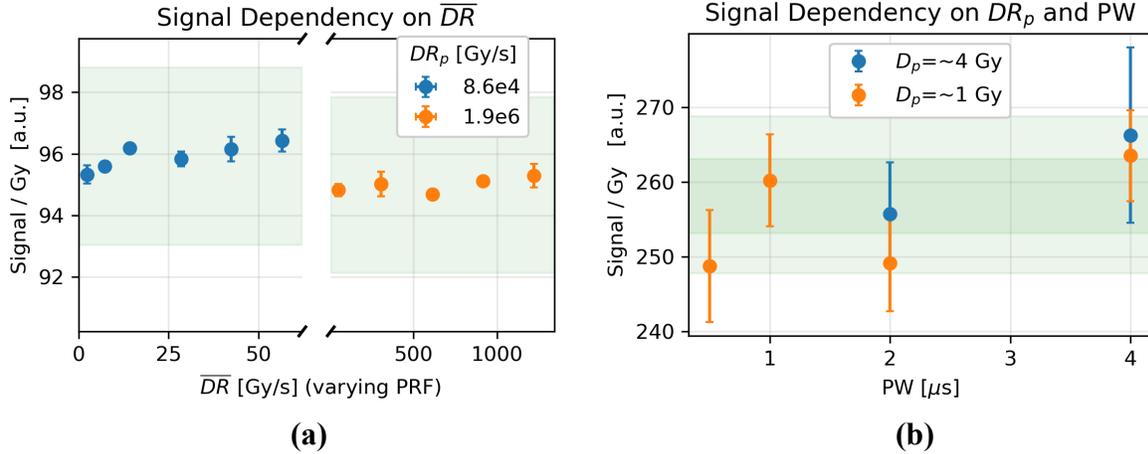

*Figure 5. a)* Dose-normalized signal change as $\overline{DR}$ was varied by changing the PRF at two values of $DR_p$. *b)* PW was varied alongside $DR_p$ to maintain equivalent values of $D_p$: $4 \pm 0.12$ Gy and $1 \pm 0.02$ Gy. The green shaded region indicates a $\pm 3\%$ variance from the mean.

**Pulse Discrimination and Pulse-by-pulse $D_p$ Measurement**

The $D_p$ recorded by the system agreed with the BCTs within ±2%, except for occasional notable outliers, where the system recorded a $D_p$ ~2–6% lower than the BCTs **(Figure 6b)**. These outliers were due to the 'split pulse' phenomenon, whereby the signal from one pulse is split, albeit largely unequally, between two adjacent integration windows of the detector **(Figure 6a)**. The assumed timing of the electron pulses relative to the system's integration windows that could have caused the observed split pulses is overlaid on **Figure 6a**. This effect can be corrected for (see Discussion for details). 13% of pulses needed a correction of 1–5%, and no pulses needed a correction of >5%. The magnitude and frequency of these split pulses are apparent in **Figure 6b** where the raw scintillator signal is low.

For the corrected pulses, as long as the sampling frequency remained higher than the Nyquist frequency ($2*PRF$), individual pulses were reliably measured without any aliasing. Sampling at exactly the Nyquist frequency did occasionally lead to aliasing, which suggests that the true sampling frequency of the system may be slightly lower than that set by the user. Sampling at double the Nyquist frequency did not reduce the occurrence rate of 'split pulses'. The average recorded $PRF$ from the system matched that of the BCTs. No differences were observed between the lower and higher tested $D_p$ and $PRF$ values.

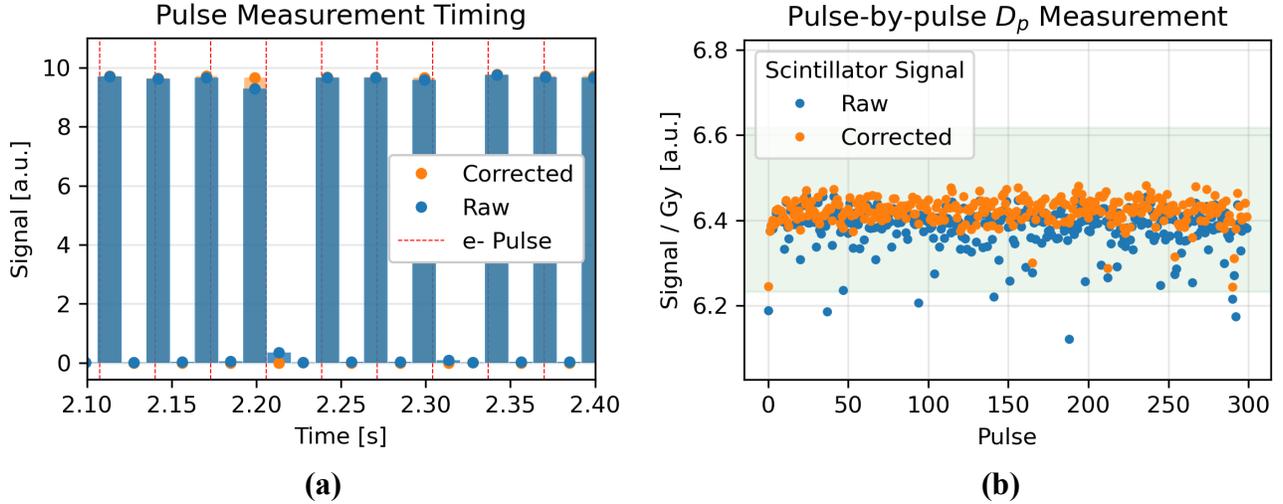

***Figure 6.*** *(a) Ten-pulse snippet from a 300-pulse, PRF = 30 Hz delivery measured at $f_s$ = 70 Hz. A correction was applied to the raw signal to correct for the occurrence of 'split pulses'. The hypothesized timing of the electron pulses delivered at 30Hz that could have caused the observed split pulses is overlaid onto the measured signal. (b) The dose-normalized scintillator signal, raw and corrected, across all 300 pulses of the delivery. The green shaded region indicates a ±3% variance from the mean.*

# Discussion

Plastic scintillators like the one studied in this work appear to be good candidates for low-LET UHDR beamlines, though only limited research into their responses at UHDRs exists. Cecchi et al.[15] utilized the Hyperscint RP100 (MedScint, Quebec City, Canada) on an UHDR x-ray tube, demonstrating $\overline{DR}$ linearity from 3–118 Gy/s. Favaudon et al.[16] used the 2-D Lynx™ detector (FIMEL, Fontenay-aux-Roses, France), demonstrating dose linearity at $DR_p$ from 0.4–3.5 MGy/s and $D_p$ *up to* 3.5 Gy, but noted that the CCD camera used to detect the scintillating light had a limited dynamic range. Poirier et al. investigated the Hyperscint RP100 on an UHDR electron beamline, demonstrating dose linearity with $D_p$ from 0.2–0.55 Gy ($DR_p$=0.04–0.11 MGy/s) and pulse counting measurements at 2.5 ms resolution. However, these pulse counting measurements suffered from a phenomenon they refer to as 'double peaks', which lead to erroneously low $D_p$ measurement on a small percentage of pulses [17]. Ashraf et al.[18] investigated the Exradin W1 (Standard Imaging, Middleton, WI), demonstrating $\overline{DR}$ independence from 50–380 Gy/s and dose linearity with $D_p$ from 0.1–1.3 Gy/s ($DR_p$=0.1–3.5 MGy/s) but noting significant radiation damage: 16% sensitivity loss per kGy. No temporally resolved measurements were reported in this investigation. Finally, Liu et al.[19] characterized the Exradin W2 (Standard Imaging), demonstrating PW dependencies and radiation damage but otherwise good performance at $D_p$ < 1.5 Gy and *PRF* < 90 Hz, but measured a nonlinear response and signal saturation at $D_p$ > 1.5 Gy and *PRF* > 90 Hz, (with $\overline{DR}$ as low as ~300 Gy/s). The commercially available

Hyperscint RP100 and Hyperscint RP200 have been previously characterized at CDRs, exhibiting excellent dosimetric responses [23-26].

In this work, we expand on the existing scintillation dosimetry literature by testing higher ranges of UHDR parameters relevant to FLASH-RT of the Hyperscint RP-FLASH scintillation dosimetry system. In line with previous publications[17], and in the absence of a formal standard established for FLASH detectors, we consider a signal variance of up to ±3% from its expected behavior as an acceptable tolerance standard.

At CDRs, an apparent limitation of the system is that $D_{iw}$ must remain within a given range, 0.028–11.56 Gy, to yield a stable signal per unit dose. Given a $D_{iw}$ value within that range, the system demonstrates excellent signal linearity with dose. The $D_{iw}$ limitation manifests as a limitation on the temporal resolution of low-dose-rate measurements. For example, at $\overline{DR}$ = 0.1 Gy/s, $IW$ must be >0.28 s to ensure $D_{iw}$ > 0.028 Gy. Similarly, as discussed below, the upper limit of $D_{iw}$ limits the maximum measurable $\overline{DR}$ and/or $D_p$.

The consequences of the $D_{iw}$ limitations manifest at UHDRs as limitations of the maximum measurable $\overline{DR}$ and/or $D_p$. Keeping $D_{iw}$ < 11.56 Gy with the system's lowest possible time resolution of $IW$ = 1 ms limits $\overline{DR}$ < 11,640 Gy/s, and thus $D_p$ < 11.56 Gy for a beam delivering at <1000Hz (avoiding multiple pulses per $IW$).

The $D_{iw}$ value limitation is likely caused by limitations in the dynamic range of the photodetector, similar to the limitation in the CCD of the Lynx system noted by Favaudon et al. [16]. Since the saturation occurs not in the scintillating material of the probe, but rather in the photodetector, the dynamic range could hypothetically be shifted, if necessary, by modifying the sensitivity of the photodetector.

Though large (up to ±13%), the signal variance across multiple days does not appear to trend with time or with accumulated dose and is therefore unlikely to be a direct cause of radiation-induced damage to or yellowing of the optical components. Since the temperature of the detector stabilized before use, it is also unlikely that temperature fluctuations contributed to this variance. The signal variance within each day was low, within ±2%. Although no conclusive explanation is apparent for the observed large variance across days, the data indicate that a 'known dose' calibration of the probe is appropriate for each new day of use, and that a subsequent variation within each day of less than ±2% can be expected. The low variance of ±0.2% across immediately subsequent measurements within each triplicate suggests that the relatively higher daily variance of less than ±2% may be attributable to positional differences in the physical setup of the detector under the beam.

The data acquired at UHDRs while changing beam parameters ($DR_p$ = 3.8e3–1.8e6 Gy/s, $\overline{DR}$ = 1.8–1,341 Gy/s, $D_p$ = 5e-3–7.68 Gy, PRF = 5-120 Hz, PW = 0.5-4 μs) indicate that these parameters, at least within the tested ranges, appear to not affect the system's dosimetric performance. However, nonlinearity was observed when $D_p$ was increased by increasing $DR_p$ via decreasing the SSD. Similar effects have been seen before on other systems[19] and were then attributed to the varying amounts of fiber exposed to the radiation field as the SSD was changed, thereby producing varying amounts of contaminating Cerenkov/fluorescence signals. However, we observed that after retaking the calibration of the probe to ensure optimal scintillation signal isolation, the trend persisted. Also, after collimating the field such that the amount of fiber exposed at each SSD was equivalent did not change the observed trend. Thus, this effect is likely not due to the varying amount of fiber exposed. Instead, this effect is more likely due to differential effects within the photodetector or in the signal processing with increasing $DR_p$.

The system's dosimetric information on a pulse-by-pulse basis showed excellent agreement with the BCTs, with the notable exception of 'split pulses', whereby the signal from one pulse was split between two adjacent IWs. This phenomenon is similar to that reported by Poirier et al. of 'double peaks', which were understood to occur when the photodetector readout coincides with the delivery of a pulse, leaving part of the pulse on the adjacent integration windows [17]. Due to the multi-channel construction and readout of the photodetector, it is partially blind to the pulse when this happens, and thus loses ~10% of the pulse's signal. The automatic processing software of the system was therefore modified to correct for 'double pulses' such that now no signal is lost, but 'split pulses' do occur, where a small fraction (<5%) of a pulse's signal is recorded in the following integration window. Split pulses do not affect the total dose reading of a pulsed beam measurement but only affect the peak heights of the individual pulses. Since no signal is lost, the effect can be corrected for as follows: the signal from each pulse is increased by the signal of the immediately following sample, and that sample's signal is decreased by the same amount.

Since the manufacturer recommends a sampling frequency of $f_s > 2 * PRF$, there is a mismatch between the delivery and sampling frequencies, leading to inconsistencies in the number of integration windows with and without delivered pulses. For example, for a $PRF$ = 30 Hz delivery measured at $f_s$ = 70 Hz, every ~3rd pulse will be followed by 2 adjacent integration windows during which no pulse arrives. This leads to an apparent periodic offset in the temporal spacing between pulses that is caused by the discrete nature of the measurement. On average, over multiple pulses, the measured $PRF$ does indeed match the delivered $PRF$. The maximum $PRF$ the system could differentiate pulses from is limited to <500 Hz by the lowest $IW$ = 1 ms, although the system was only tested in this work up to 120 Hz, the maximum $PRF$ of the FLASH Mobetron. For the pulse-by-pulse measurements of 300 pulses, a relatively low $D_p$ of 0.1 Gy was chosen to avoid delivering very large doses to the probe during a single measurement.

As opposed to matching the sample frequency to twice the $PRF$, a fixed $IW$ of 4.1 ms ($f_s$ = 244 Hz) was set for all UHDR measurements. This is due in part to the aforementioned usage of a slightly higher sampling frequency than the Nyquist frequency, but also to the higher variance in the delivery $PRF$ of the FLASH Mobetron at its maximum output.

This study was limited in part by the output limitations of the FLASH Mobetron. The highest $PRF$ tested was 120 Hz, whereas the scintillation system could theoretically measure a beam $PRF$ of 500 Hz without aliasing effects. The tested $D_p$ and $DR_p$ were limited to 7.68 Gy and 1.8e6 Gy/s, respectively. Measuring the dependency of the system on $DR_p$ and $PW$ as $D_p$ was held constant was limited by the slight variance in the imperfectly constant $D_p$ over the tested ranges, and by having only 2 data points (2 $PW$ values) for which a $D_p$ of 4 Gy could be tested. The observed reduction in signal per unit dose as $DR_p$ was increased should be further investigated over a wider range of $DR_p$. Additionally, beam energy dependency of the system was not tested.

**Conclusions**

We performed a comprehensive investigation of the dosimetric performance of the scintillation system across the full range of parameters possible on the FLASH Mobetron. The system was linear with dose at both CDRs and UHDRs and showed no dependence on any beam parameters throughout the tested ranges, apart from a 6% signal decrease when increasing the $DR_p$ through reduction of SSD and the limits of the dynamic range of the photodetector, which require that the dose per integration window of the photodetector remain within 0.028–11.64 Gy. Individual pulses could be properly resolved and $D_p$ measured at 1 ms time resolution and, after applying a simple post-measurement correction for an effect coined 'split pulses', agreed with the BCTs within ±2%. Daily variance of the signal remained lower than ±2%, but up to ±13% variance across days suggests the necessity of a known-dose calibration before each day of use. This study demonstrates the first to-date scintillator dosimetry system capable of providing online and millisecond-resolved dosimetric measurements over the entire dynamic range of CDRs and UHDRs from typical electron linacs, marking an important advancement in UHDR dosimetry and offering diverse applications to FLASH-RT and related fields.